\documentclass[onecolumn]{emulateapj}
\usepackage{graphicx}
\usepackage{lineno}
\usepackage{longtable}
\usepackage{amsmath}
\usepackage{xspace}
\usepackage[colorlinks]{hyperref}
\usepackage[colorinlistoftodos]{todonotes}

\usepackage{natbib}

\newcommand{\Fermi}{\emph{Fermi}\xspace}
\newcommand{\fermi}{\emph{Fermi}\xspace}
\def\de{$^{\circ}$}

\setkeys{Gin}{draft=false}

\shortauthors{{Winter {\it et al.}}}
\shorttitle{{Long-duration Gamma-ray Solar Flares}}

\begin{document}


\title{A Statistical Study to Determine the Origin of Long-Duration Gamma-ray Flares}  

\author{L.M. Winter}
\affil{Space Science and Applications, Los Alamos National Laboratory, Los Alamos, NM, 87545, USA}
\email{lmwinter@lanl.gov}

\author{V. Bernstein}
\affil{Ann and H.J. Smead Aerospace Engineering Sciences, University of Colorado Boulder, Boulder, CO, 80309, USA}


\author{N. Omodei}
\affil{W. W. Hansen Experimental Physics Laboratory, Kavli Institute for Particle Astrophysics and Cosmology, Department of Physics and SLAC National Accelerator Laboratory, Stanford University, Stanford, CA, 94305, USA}

\author{M. Pesce-Rollins}
\affil{Istituto Nazionale di Fisica Nucleare, Sezione di Pisa,I-56127 Pisa, Italy}


\begin{abstract}
Two scenarios have been proposed to account for sustained $\ge 30$\,MeV gamma-ray emission in solar flares:  (1) prolonged particle acceleration/trapping involving large-scale magnetic loops at the flare site, and (2) precipitation of high-energy ($>$ 300 MeV) protons accelerated at coronal/interplanetary shock waves. To determine which of these scenarios is more likely, we examine the associated soft X-ray flares, coronal mass ejections (CMEs), and solar energetic proton events (SEPs) for: (a) the long-duration  gamma-ray flares (LDGRFs) observed by the Large Area Telescope (LAT) on \Fermi, and (b) delayed and/or spatially-extended high-energy gamma-ray flares observed by the Gamma-ray Spectrometer on the Solar Maximum Mission, the Gamma-1 telescope on the Gamma satellite, and the Energetic Gamma-Ray Experiment Telescope on the Compton Gamma-Ray Observatory.  For the \Fermi data set of 11  LDGRFs with $>$100 MeV emission lasting for $\ge \sim 2$ hours, we search for associations and reverse associations between LDGRFs, X-ray flares, CMEs, and SEPs, i.e., beginning with the gamma-ray flares and also, in turn, with X-class soft X-ray flares, fast ($\ge$ 1500 km s$^{-1}$) and wide CMEs, and intense (peak flux $\ge 2.67 \times 10^{-3}$ protons cm$^{-2}$ s$^{-1}$ sr$^{-1}$, with peak to background ratio $>$ 1.38) $>$ 300 MeV SEPs at 1 A.U. While LDGRFs tend to be associated with bright X-class flares, we find that only 1/3 of the X-class flares during the time of \Fermi monitoring coincide with an LDGRF. However, nearly all fast, wide CMEs are associated with an LDGRF. These preliminary association analyses favor the proton precipitation scenario, although there is a prominent counter-example of a potentially magnetically well-connected solar eruption with $>$ 100 MeV emission for $\sim 10$ h for which the near-Earth $>$ 300 MeV proton intensity did not rise above background.       
\end{abstract}

\section{Introduction}

Gamma-ray solar flares provide the opportunity to examine particle acceleration mechanisms associated with solar activity. In particular, long-duration gamma-ray flares present a challenge to models since they exhibit delayed emission lasting several hours. This suggests an extended acceleration phase inconsistent with an impulsive flare origin. { Further, it is difficult to distinguish between gamma-ray continuum emission produced by bremsstrahlung (electrons) and pion decay (hadrons)}. Therefore, the origin of long-duration gamma-ray flares is not well understood. 


The 1982 June 3 event provided the first clear evidence for pion-decay emission - which requires the acceleration of $>$ 300 MeV protons for pion production - in solar flares \citep{1985ICRC....4..146F}. Accelerated protons interact with the chromosphere and can produce pions which may subsequently decay to emit electrons, positrons, and/or gamma-rays. We would expect the pion-decay process to occur only during the impulsive phase of a flare, but for this event, gamma-ray emission was observed distinctly during the flare's impulsive phase and then in an electron (i.e., hard X-rays $< 10$ MeV) deficient phase that began $\sim 2$ min later. \citet{1986AdSpR...6..115F} reported that $>70$\% of the pions in the 1982 June 3 event were produced in the extended ($\sim 20$ min) delayed phase. Subsequently, \citet{1993A&AS...97..349K} reported a prolonged ($\sim $8 h) phase of $>50$ MeV gamma-ray emission in a flare on 1991 June 11 that they modeled in terms of a proton-generated $\pi^{0}$ spectrum.
	
The delayed/prolonged high-energy emission in these flares and others \citep{2000SSRv...93..581R,2009RAA.....9...11C} observed by the Gamma-ray Spectrometer (GRS) \citep{1980SoPh...65...15F} on the Solar Maximum Mission (SMM), the Gamma-1 telescope \citep{1988SSRv...49..111A} on the Gamma satellite, and the Energetic Gamma-Ray Experiment Telescope (EGRET) \citep{1988SSRv...49...69K} on the Compton Gamma-Ray Observatory (CGRO) confronts the traditional view that the flare impulsive phase encompasses the most energetic flare emissions \citep{2011SSRv..158....5H,2011SSRv..159...19F}. As \citet{2000SSRv...93..581R} put it, ``The intuitive notion that energy degrades in form over time seems not to hold in these circumstances."  Ryan introduced the term ``long-duration gamma-ray flares" or LDGRFs, which we will use here.
	
The first explanation proposed for sustained pion-decay gamma-ray emission in flares linked such events to the solar energetic protons (SEPs) observed in space following flares. \citet{1985ICRC....4..146F} noted that both the SEP event associated with the 1982 June 3 event flare \citep{1985ApJ...290L..67M} and the delayed gamma-ray phase observed by SMM GRS exhibited unusually hard spectra. Subsequently, \citet{1987ApJ...316L..41R} proposed that the SEPs observed in space and those responsible for the delayed high-energy gamma-ray emission in this event had the same origin, a coronal Type II shock \citep{1963ARA&A...1..291W,1987ApJS...63..721M}. Similar suggestions were made by \citet{1993ApJ...409L..69V} and \citet{1993ICRC....3...91C} to account for spatially-extended gamma-ray emission in the 1989 September 29 eruptive flare that originated behind the west limb of the Sun.
	
Various flare-centric alternatives to the coronal shock origin of temporally and/or spatially extended pion-dominated gamma-ray events include:  impulsive acceleration of particles onto large loops (up to $10^5$ km) with trapping and variable precipitation \citep{1990SoPh..125...91G,1992ApJ...389..739M}; continuous acceleration and diffusion in a turbulent storage loop \citep{1991ApJ...368..316R}; multiple episodes of particle acceleration due to reconnection in a current sheet in the wake of a CME \citep{1993ICRC....3..111A,1996SoPh..166..107A}; episodal acceleration and subsequent trapping \citep{1996AIPC..374..225M}. For reviews of observations and models of LGDRFs, see \citet{1995ARA&A..33..239H}, \citet{2000SSRv...93..581R}, and \citet{2009RAA.....9...11C}.
	
Recently, new observations of LDGRFs by the Large Area Telescope (LAT; \citealt{2009ApJ...697.1071A}) on the \Fermi spacecraft have provided support for the coronal shock origin of the $> 300$ MeV protons required for the production of pions.  \citet{2015ApJ...805L..15P} and \citet{2017ApJ...835..219A} interpreted the observation of three $> 100$ MeV gamma-ray producing events associated with behind-the-limb flares from occulted active regions in terms of coronal mass ejection (CME) driven shock acceleration of $> 300$ MeV protons which then precipitated to the photosphere to create pions. 
	
In this study of the origin of sustained high-energy gamma-ray emission in solar flares, we follow up on the results of these two LAT papers and ask whether LDGRFs are more likely to be associated with big flares (favoring scenarios involving proton acceleration via magnetic reconnection in the flaring region) or fast CMEs and high-energy SEPs (supporting the shock scenario). In Section 2, we describe the various data sets used. In the analysis of these data in Sections 3-5, we first discuss the GOES X-ray flare (Section 3), CME (Section 4), and SEP (Section 5) properties of the 29 $> 100$ MeV flares observed to date by \Fermi LAT, focusing on the 11 events that have nominal durations of $\sim 2$ h or more. We then include an overview of the flare, CME, and SEP events of the early events observed by SMM, Gamma, and Compton GRO. For each of the three phenomena considered for association with the LDGRFs -- flares, CMEs, SEP events -- we make a reverse search for outstanding cases of each of these phenomena to determine the fractions of these events that were accompanied by sustained $>$ 100 MeV gamma-ray emission. Our results are summarized and discussed in Section 6.

\section{Observations}
To establish possible connections between gamma-ray flares, X-ray flares, CMEs, and SEPs, several data sources were used. 
The LAT is a wide field of view (FOV), imaging telescope for high-energy gamma-rays, designed to cover an energy range from 20 MeV up to more than 300 GeV~\citep{2009ApJ...697.1071A}. The instrument consists of a precision tracker with silicon strip detectors above a cesium-iodide calorimeter. Both are enclosed in the plastic scintillators of the Anti-Coincidence Detector that provides charged-particle tagging for background rejection. 
The LAT monitors the entire sky every two orbits, or about every three hours, and observes the Sun for $\sim 20 - 40$ contiguous minutes in that time. The LAT team has created an automated data analysis
pipeline, the LAT \texttt{SunMonitor}, to monitor the high-energy gamma-ray flux from the Sun throughout the \fermi mission. The time intervals during which the analysis is run are the intervals in which the Sun is less than 60\de~off-axis for the LAT.
In this way, each interval corresponds to the maximum time with continuous Sun exposure, and the duration of these intervals varies as the Sun advances along the ecliptic and as the orbit of \Fermi precesses. { Since the results from the LAT \texttt{SunMonitor} are run on fixed time intervals, the light curves produced are not derived from subdivided, shorter time intervals. However, in each time window, an unbinned likelihood analysis of the LAT data is performed with the \texttt{gtlike} program distributed with the \Fermi \texttt{ScienceTools}\footnote{Available from the \Fermi Science Support Center \url{http://fermi.gsfc.nasa.gov/ssc/}}, for which the time interval can be subdivided as necessary.} The analysis is based on Pass 8 \texttt{SOURCE} class events with energies above 100 MeV and coming from { within} 12\de~of the Sun and within 100\de~from the local zenith (to reduce contamination from the Earth limb).  Further details on the analysis procedure used can be found in~\citep{2014ApJ...787...15A, 2014ApJ...789...20A}. The likelihood ratio test and the associated test statistic TS~\citep[]{Mattox:96} are used to estimate the significance of the LAT detection. The TS is approximately the square of the significance. \Fermi LAT gamma-ray flare detections, as listed in Table~1, are  those cases where there is a rise in flux along with a statistically significant TS value greater than 25 (i.e. $\sigma>$ 5). The LAT fluence is calculated by integrating over a linear interpolation of the measured flux over all pointing with a flare detection in the \texttt{SunMonitor}. For flares detected in only a single pointing, the LAT fluence is computed as a product of the flux and the time interval of the detection. Example \Fermi LAT light curves for the flares we discuss are published in previous studies including \citet{2014ApJ...787...15A,2017ApJ...835..219A}.



The NOAA Geostationary Operational Environmental Satellites (GOES) provide continuous monitoring of the solar X-ray and energetic particles reaching Earth. Over the same time period as the \Fermi LAT observations (2010 -- 2016), we analyzed the X-ray flare and solar energetic particle properties from the GOES-13 and GOES-15 satellites. X-ray measurements investigated include the peak X-ray flare flux and start time in the 1--8\,\AA~X-rays from the X-Ray Sensor (XRS) from the NOAA X-ray Flare List\footnote{The NOAA X-ray Flare List is available at \url{http://www.ngdc.noaa.gov/stp/satellite/goes/dataaccess.html}}. The official NOAA criteria for defining the start of a flare is when four consecutive one-minute 1--8\,\AA~flux measurements meet the following conditions: (1) Each of the four consecutive one-minute flux measurements $> 10^{-7}$\,W\,m$^{-2}$, (2) each consecutive measurement has a higher flux than the previous measurement, and (3) the last of the four X-ray measurements is $> 1.4\times$ the measurement from three minutes earlier. We also measured the maximum ratio of the 0.5-4\,\AA/1-8\,\AA\,GOES flux, $R_{\rm max}$, during each flare as a measure of the thermal energy of the flare, following the procedure in 
\citet{2015SpWea..13..286W}. The flare location was obtained from either the NOAA flare list\footnote{ The NOAA Flare List was obtained from \url{https://www.ngdc.noaa.gov/stp/solar/solarflares.html\#xray}.} or the {\it Solar and Heliospheric Observatory} (SOHO) {\it Large Angle and Spectrometric Coronagraph} (LASCO) Coordinated Data Analysis Workshops (CDAW) CME catalog\footnote{A description of the LASCO CDAW catalog fields is available at \url{http://cdaw.gsfc.nasa.gov/CME_list/catalog_description.htm}}, and is based on H$\alpha$ observations.

SEP properties were investigated from the Space Environment Monitor (SEM), including 5-min averages in the integral energy band of $> 10$\,MeV. The NOAA definition for an SEP event is a $> 10$\,MeV flux above 10\,pfu { (pfu $=$ protons s$^{-1}$ cm$^{-2}$ sr$^{-1}$)}. We also analyzed the GOES SEM high energy proton and alpha detector (HEPAD) observations in differential energy channels of 370 - 480 MeV (P8), 480 - 640 MeV (P9), 640 - 850 MeV (P10). Changes in the SEP profiles in each of these channels were consistent (i.e., similar peak times, profile shape), so we focused on the P8 channel, which is closest in energy to that of protons involved in the pion-decay process that produces gamma-ray emission in the \Fermi LAT band. We searched for SEP events over the entire time period from 2011--mid-2015, detecting all local maxima above the background level. The peak flux in Table~1 represents the highest flux measured in a 1-day interval following the onset of the associated X-ray flare.

CME properties, including on-set time, angular width, and linear speed were obtained from the SOHO LASCO CDAW catalog. The LASCO linear speed measurements result from fitting a line through the CME height-time data (i.e., linear fits to coronagraph plane of sky speeds). The LASCO width measurements are sky-plane widths (i.e., position angle extent of CMEs which are upper limits on the true CME width). The majority of CMEs associated with gamma-ray flares are halo or partial halo CMEs, with a few cases of wide angle ($>100$ degrees), non-halo CMEs.

We additionally examined past long duration gamma-ray flares from earlier studies to test for X-ray flare, CME, and SEP connections. We focused on 13 gamma-ray producing events observed prior to the launch of \Fermi LAT compiled by \citet{2000SSRv...93..581R}. These gamma-ray events were observed between the years of 1982 and 1991, and gamma-ray measurements were obtained by GRS on SMM and EGRET on CGRO. This sample of pre-\Fermi LAT events includes a range of gamma-ray flare durations, the shortest of which was attributed to the 1990-05-24 event which lasted 8.3 minutes (\citealt{1994SoPh..155..149K}; \citealt{1997ApJ...479..997D}), while the longest duration was attributed to the 1991-06-11 event which lasted 8.3 hours (\citealt{1993A&AS...97..349K}; \citealt{1996AIPC..374..219R}; \citealt{1996A&AS..120C.299S}). Three out of the 13 pre-\Fermi LAT events lasted longer than 1 hour, including the previously mentioned 1991-06-11 event, the 1991-06-04 event which had a duration of 2.8 hours (\citealt{1996A&AS..120C.299S}; \citealt{1997ApJ...490..883M}), and the 1991-06-15 event which had a duration of 1.4 hours (\citealt{1994AIPC..294..130A}; \citealt{1996AIPC..374..219R}). In all cases, these events exhibit a two-component intensity-time profile at energies $> 50$\,MeV, which leads to their classifications as LDGRFs. The emission profiles exhibit a short burst followed by an exponential decay.

We cross-examined these pre-\Fermi LAT events with CME observations from the SOLWIND white light coronograph \citep{1980ApJ...237L..99S} on the Air Force Space Test Program satellite P78-1 and from the SMM coronograph \citep{1980SoPh...65...91M}. SOLWIND captured images from the solar corona from March, 1979, through September, 1985, while the SMM obtained CME information from March, 1980, through November, 1989. We additionally analyzed X-ray flux and solar energetic particle observations between 1986 and 1991 from the GOES-06 and GOES-07 satellites. We analyzed the GOES-06 SEM HEPAD observations in the P8, P9, and P10 differential energy channels, as described above. For the GOES-07 observations, we analyzed data from the Energetic Particle Sensor (EPS) instrument, time-averaged over 1 minute, in differential energy channels of 84 - 200 MeV (P6) and 110 - 500 MeV (P7).

Table~\ref{tbl-fermi} includes details of the \Fermi LAT gamma-ray, CME, X-ray, and energetic proton properties discussed above. The \Fermi LAT gamma-ray list originally included all solar gamma-ray flares with $\sigma>$ 5. Through our analysis, we also looked for fast CMEs (linear speed $> 1500$\,km\,s$^{-1}$), X-class X-ray flares, or SEP events in the GOES HEPAD observations without a \Fermi LAT detection. Fast CMEs and SEPs without accompanying \Fermi LAT detections are included in Table~\ref{tbl-fermi}. Table~\ref{tbl-nofermi} lists the properties of the GOES X-class flares without detected gamma-ray emission. Table~\ref{tbl-prefermi} includes timing and duration details for gamma-ray events detected prior to the launch of \Fermi LAT along with corresponding CME, X-ray, and SEP properties. In this paper, we focus on the associated CME, X-ray flare, and SEP properties of both the \Fermi LAT and pre-\Fermi gamma ray events and the implications of their properties with respect to the origin of LDGRFs. 


\section{Lack of Strong Connection with X-ray Flares}
\subsection{Properties of Associated GOES X-ray Flares with \Fermi LAT Flares}
{ X-ray sources are created when energetic particles accelerated from the corona to the chromosphere heat and interact with dense plasmas in the chromosphere. Energy can be released as X-rays when magnetic field loops become twisted or when these plasmas are heated and pushed upward \citep{1968ApJ...153L..59N}. Hard X-ray emission peaks are generally considered to be impulsive and follow the rapid crossing of bright flare loops, while SXR emissions are considered to be more thermally dominated and can be delayed by less than a few minutes \citep{1970ApJ...162.1003K}. Further, hard X-ray emission of up to hundreds of keV suggest the acceleration of MeV electrons producing bremsstrahlung radiation in dense regions of the corona and chromosphere. At higher energies, the 1-10 MeV gamma-ray lines are produced when accelerating ions, excited through collisions, release energy and become de-excited. $> 100$\,MeV continuum radiation is produced when $> 300$\,MeV protons collide with $> 800$\,MeV alpha particles to create neutral and charged pions which decay to produce gamma-rays \citep{1987ApJS...63..721M}. Neutral pions decay into 67.5 MeV gamma-ray pairs. Charged pions decay into neutrinos as well as electrons and positrons which emit MeV bremsstrahlung gamma-rays. Gamma-ray emission has an impulsive phase lasting up to a few minutes with a profile typically consistent with hard X-ray emission. Long-duration gamma-ray flares have presented with extended ($> 1$\, hour) emission with an exponential decay which have possible electron bremsstrahlung and pion-decay components \citep{2001A&A...378.1046R}. Pion-decay processes are considered to be more likely than electron bremsstrahlung because pion-decay processes require a smaller energy input, and $> 100$\,MeV long-duration gamma-rays have a total energy that is hundreds of times smaller than the hard X-rays during the impulsive phase \citep{2014ApJ...787...15A}.

If the gamma-ray emission from the long-duration \Fermi flares is continuum emission associated with the X-ray flare, we would expect to find a correlation between the X-ray and gamma-ray duration and/or flux. We tested this by comparing the \Fermi LAT  flux of the detected solar flares with the GOES 1--8\,\AA\,peak flux and the ratio of the maximum in the GOES 0.5-4\,\AA/1-8\,\AA\,flux during the flare. We excluded known cases of behind the limb flares, like the strong gamma-ray flare on September 1, 2014, since direct X-ray observations of the associated flare are not available.    }

\begin{figure}
\includegraphics[width=0.95\textwidth]{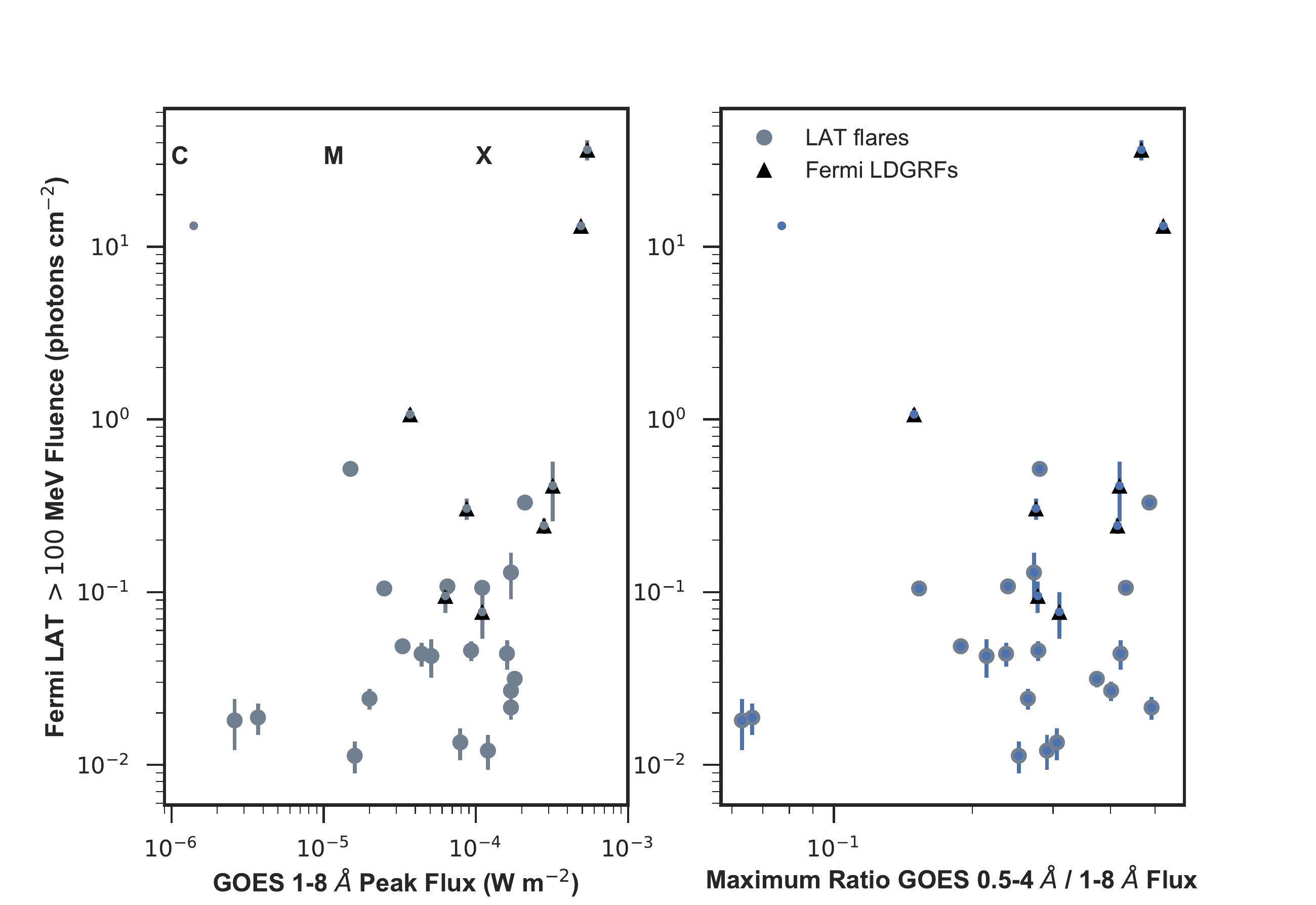}
\caption{\Fermi LAT fluence (see Table 1) compared to the associated X-ray flare class (left) and $R_{max}$ (right). Long duration gamma-ray flares (with duration $>$ 2 hours), are indicated with triangles, while shorter duration gamma-ray flares are shown with circles. There is no evidence of a correlation between the peak flux or flare temperature diagnostic ($R_{max}$) and the gamma-ray emission.
}\label{fig-xray}
\end{figure}

Figure~\ref{fig-xray} shows the \Fermi LAT $> 100$\,MeV fluence related to the X-ray peak flux and flare temperature diagnostic, $R_{max}$. \Fermi LAT fluence is measured by the \Fermi-LAT \texttt{SunMonitor} between 100 MeV and 10 GeV during times when the Sun is positioned $\leq 60$\de~from the axis of sight of LAT such that gaps in monitoring exist when LAT is directed away from the Sun. For the longest duration gamma-ray flares, with durations over 2 hours, we performed a linear regression analysis to test for associations between \Fermi LAT fluence and X-ray flare properties. We find, as shown in the figure, that there is no statistical correlation between the \Fermi LAT fluence and the X-ray diagnostics. { The Kendall's $\tau$ and $p$-values are 0.43 and 0.14 for the comparison with X-ray peak flux and 0.29 and 0.32 for the comparison with $R_{max}$.}

Since the \Fermi LAT fluence is integrated over time periods when the satellite is directed towards the Sun, there are gaps in monitoring throughout the entire flare corresponding to when the Sun is not in the field of view of \Fermi LAT. This could account in part for the lack of correlation with the X-ray, as the reported \Fermi LAT fluence may be representative of a lower limit of the integrated flux. However, we note that for any X-ray flare class or temperature (indicated by $R_{max}$), there is a large range of \Fermi fluence. For instance, the highest measured gamma-ray fluence flares, the 6 flares with $F_{LAT} > 1\times10^{-4}$\,photons\,cm$^{-2}$\,s$^{-1}$, have an average X-ray class of X1.5 compared to an average class of M9.5 for gamma-ray flares with lower fluence. The X-ray temperature diagnostics are very similar, corresponding to an average and standard deviation for the higher gamma-ray fluence flares as $<R_{max}> = 0.29 \pm 0.16$ and $<R_{max}> = 0.26 \pm 0.13$ for the lower fluence gamma-ray flares. The large standard deviations show that the high fluence gamma-ray flares have similar X-ray flare properties to the lower fluence gamma-ray events. Therefore, there is no significant correlation seen between the X-ray and gamma-ray flares. These statistical results suggest a lack of support for the scenario in which extended gamma-ray emission can be attributed to magnetic reconnection in the flare loop, which is associated with the production of X-ray flares.

\subsection{Properties of Associated GOES X-ray Flares with Pre-Fermi Gamma-ray Flares}
We examined the properties of X-ray flares associated close in time with the 13 pre-\Fermi gamma-ray events, detected with GRS on SMM, Gamma-1, and EGRET on CGRO. Since these flares were not selected in a uniform survey, we can not comment on whether there were strong pre-\Fermi X-ray flares with no gamma-ray emission as we will for our reverse study analyses for X-ray flares, CMEs, and SEPs in search of accompanying \Fermi LAT gamma-ray emission. Also, the gamma-ray detection methods vary from, for example, measuring nuclear capture gamma-ray lines with SMM's GRS instrument to diffuse gamma-ray emission above $>50$\,MeV and neutron capture lines with CGRO instruments. This means that the definitions of the intensity and duration of the LDGRFs are not comparable. However, we can note that all of the pre-\Fermi gamma-ray flares are associated with strong X-class flares. These LDGRFs were detected in earlier, more active solar cycles, accounting for the much higher X-class of the majority of pre-\Fermi LDGFs. While only one \Fermi LDGRF is above X5 class, 10/13 of the pre-\Fermi LDGRFs are $>$X5 class and 7/13 are $>$X10 class.

The 1989-09-29 gamma-ray event is particularly notable, discovered by \citet{1993ApJ...409L..69V} as the first behind-the-limb gamma-ray flare. The X-ray flare was the strongest associated with gamma-ray emission (X15), but located on the eastern limb (N35E69). The gamma-ray emission from this event was also shown to be spatially extended by 30$^{\circ}$, which is unexpected if the gamma-ray emission were to originate in the magnetic footpoints of the X-ray flare. 

The relative timing of the X-ray and gamma-ray emission for the pre-\Fermi LDGRFs tends to be near simultaneous (i.e., within minutes). Two exceptions include the September 29, 1989 and June 15, 1991 events. For the 09/1989 event the H$\alpha$ flare location is towards the eastern limb. The X-ray emission begins $\sim 2$ hours before the gamma-ray emission is detected. 
For the June 15, 1991 event, located at N33W69, the X-ray flare leads the gamma-ray emission by 2 hours.


\subsection{Strong GOES X-ray Flares with No Fermi LAT Detection}
 The strongest evidence for a lack of correlation between the GOES X-ray flare and the gamma-ray flare emerges from examining the NOAA-identified GOES X-ray flares without \Fermi detections. Using the NOAA SWPC flare list, we find that 45 X-class flares occurred during the time period of our study. However, only 15 or one-third of these flares are associated with \Fermi LAT gamma-ray detections. Our identification of \Fermi LDGRFs is as gamma-ray flares detected in multiple orbits with LAT. 
 
\begin{figure}
\includegraphics[width=0.45\textwidth]{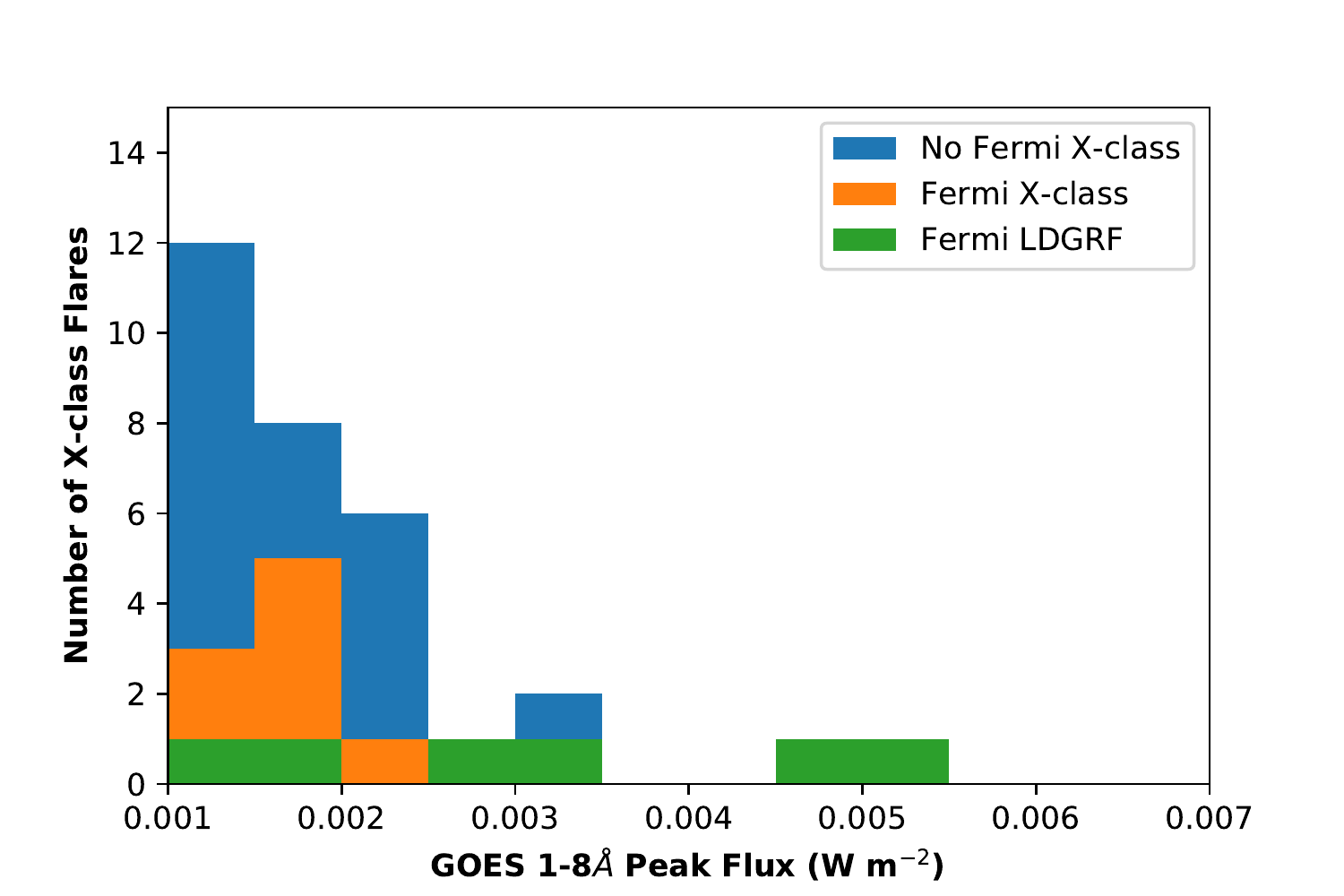}
\includegraphics[width=0.45\textwidth]{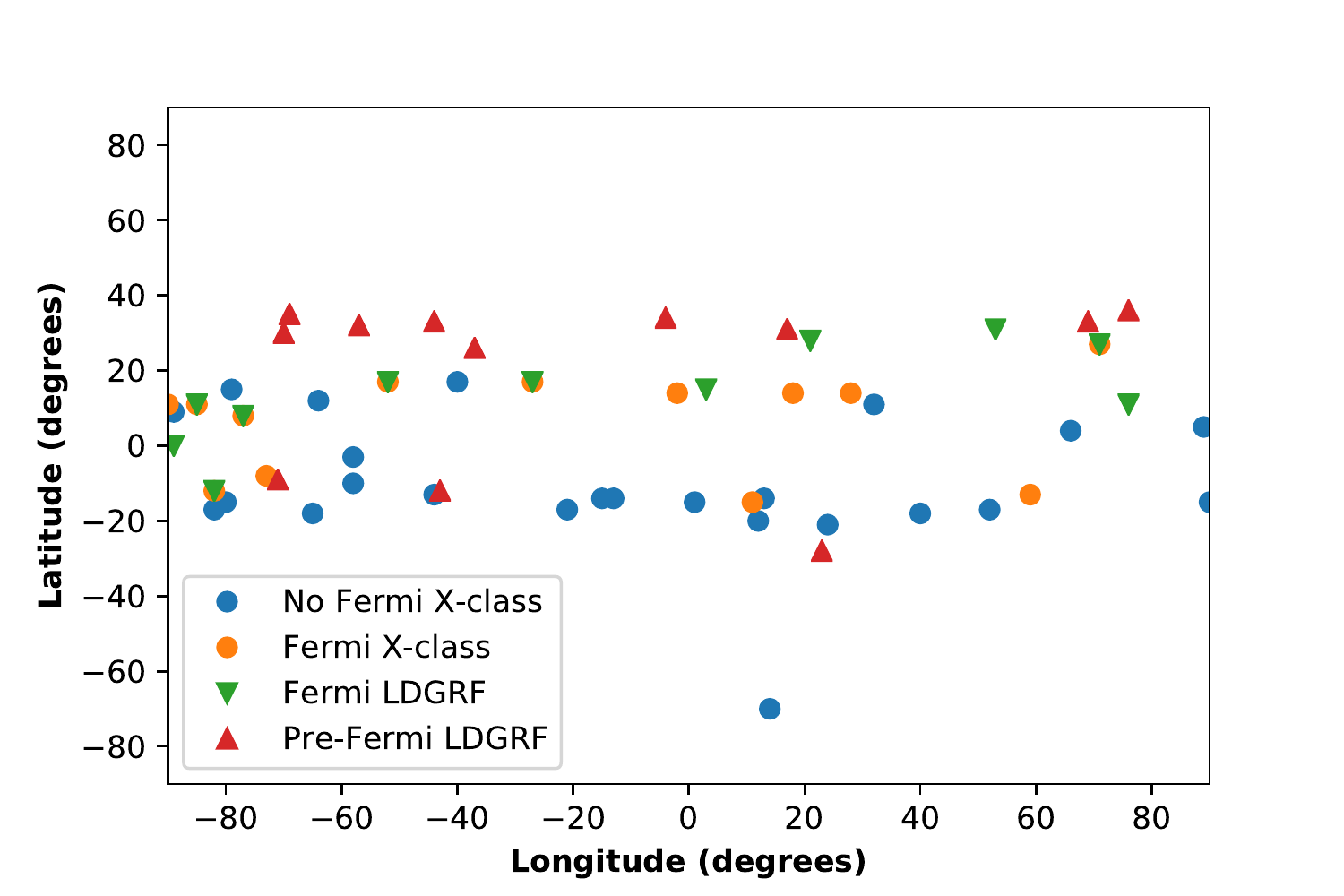}
\caption{Comparison of the flux and location of GOES X-class flares with and without \Fermi LAT detections and \Fermi LDGRFs. (Left) Distribution of the NOAA X-classification shows that flares with (orange) and without (blue) gamma-ray detections do not have significant differences in their X-ray flux. Only half of the LDGRFs are X-class, showing that the brightest X-ray flares are not necessarily the brightest gamma-ray sources. (Right) The flare location for X-class flares (blue) spans the active regions of the Sun during this time period. However, those with \Fermi gamma-ray emission are preferentially associated with locations in the North (orange and green). This is also true of the pre-\Fermi LDGRFs (red).
}\label{fig-xclassldgrfs}

\end{figure}

 In Table~\ref{tbl-nofermi}, the associated CME, SEP, and X-ray flare properties are shown for the 30 X-class flares without \Fermi detections. The distributions of peak flux for these sources are included in Figure~\ref{fig-xclassldgrfs} (left), compared to the X-class flares with \Fermi detections and those associated with LDGRFs. The median classification for non-\Fermi flares is X1.6, while it is X1.7 for those with detections. Likewise, there is no difference in the temperature of the flares with and without gamma-ray emission, indicated by both populations having median $R_{max} = 0.41$. Further, the \Fermi flares associated with LDGRFs have a median peak flux (X1.1) and temperature ($R_{max} = 0.27$) lower than the typical X-class flares occurring over the same time period (2011-2015).

Among the brightest X-ray flares, there were six X3 or higher X-ray flares. The brightest was an X6.9 flare at N14W69, on 2011-08-09. LAT detected only impulsive emission for this flare, using the LAT Low Energy (LLE) analysis \citep{0004-637X-787-1-15}. Additionally, an X3.3 flare on 2013-11-05 at S12E47 and an X3.1 flare on 2014-10-24 at S22W21 had no associated \Fermi LAT detection. Only half of the brightest X-class flares during this time period are associated with \Fermi LAT gamma-ray emission.

The locations of the flare sites are different between flares with and without \Fermi detections. This is shown in Figure~\ref{fig-xclassldgrfs} (right). Here, the median location for X-class flares with \Fermi detections (orange) is N11E27. For those X-class flares without a \Fermi detection (blue) the median location is S14E14. Interestingly, the location for the pre-\Fermi LDGRFs is also in the northern-eastern hemisphere (N31E37) as are the \Fermi LDGRFs (N15E27). It is unclear why there might be a north-eastern hemisphere preference for detecting associations with LDGRFs, but it likely has to do with the viewing angle between Earth and the site on the solar surface where the gamma-ray emission is produced.

Therefore, we can make two conclusions based upon the analysis of X-class flares with and without associated gamma-ray emission. The first is that the LDGRFs are not associated with all bright X-ray flares. The second is that the flare location likely plays a role in whether the gamma-ray emission is detected. LDGRFs, from pre-\Fermi and \Fermi are predominately in the north-eastern hemisphere. Previous studies of gamma-ray solar flares found that emission is detected more frequently near the limb and that significant center-to-limb variations in ratios of soft to hard flux are seen in flares \citep{1987ApJ...322.1010V}. Our results support the evidence that the gamma-ray emission is not isotropic and potentially beamed. Not only is it more likely to detect flares on the limb, but our view from Earth also preferentially detects gamma-ray emission towards the North-East limb.



\newpage
\section{Correlation with CME properties}
{ Continuous acceleration of particles by a CME shock front has been proposed to produce extended gamma-ray emission \citep{2001A&A...378.1046R}. CME shocks are capable of accelerating particles to $>300$\,MeV energies necessary for gamma-ray production. However, gamma-ray emission requires high-density source regions, thus CME-accelerated protons and electrons must be transported along magnetic field lines from low-density regions behind the CME to high-density regions below the corona \citep{2013SSRv..175...53R}. In this section, we discuss the evidence for connections between CME properties and gamma-ray flares.
}

\subsection{Properties of CMEs Near in Time with Fermi LAT Flares}
Recent \Fermi LAT studies show evidence supporting the origin of long-duration gamma-ray flares from acceleration in CME shocks, particularly through the gamma-ray detection associated with behind-the-limb flares  
\citep{2015ApJ...805L..15P}. To search for connections between the larger sample of gamma-ray flares and CMEs, we compared the timing of the gamma-ray flares with CMEs in the SOHO LASCO CDAW catalog. Table~\ref{tbl-fermi} shows that all of the long duration gamma-ray events are associated with CMEs. There are only 5 events not associated with an observed CME, and all are shorter duration and lower gamma-ray fluence events. The \Fermi LAT flares are associated, on average, with wide angle CMEs (i.e., the majority are halo CMEs). These CMEs are also fast, with an average linear speed is 1482\,km\,s$^{-1}$ and a standard deviation of 655\,km\,s$^{-1}$. 

\begin{figure}
\includegraphics[width=0.95\textwidth]{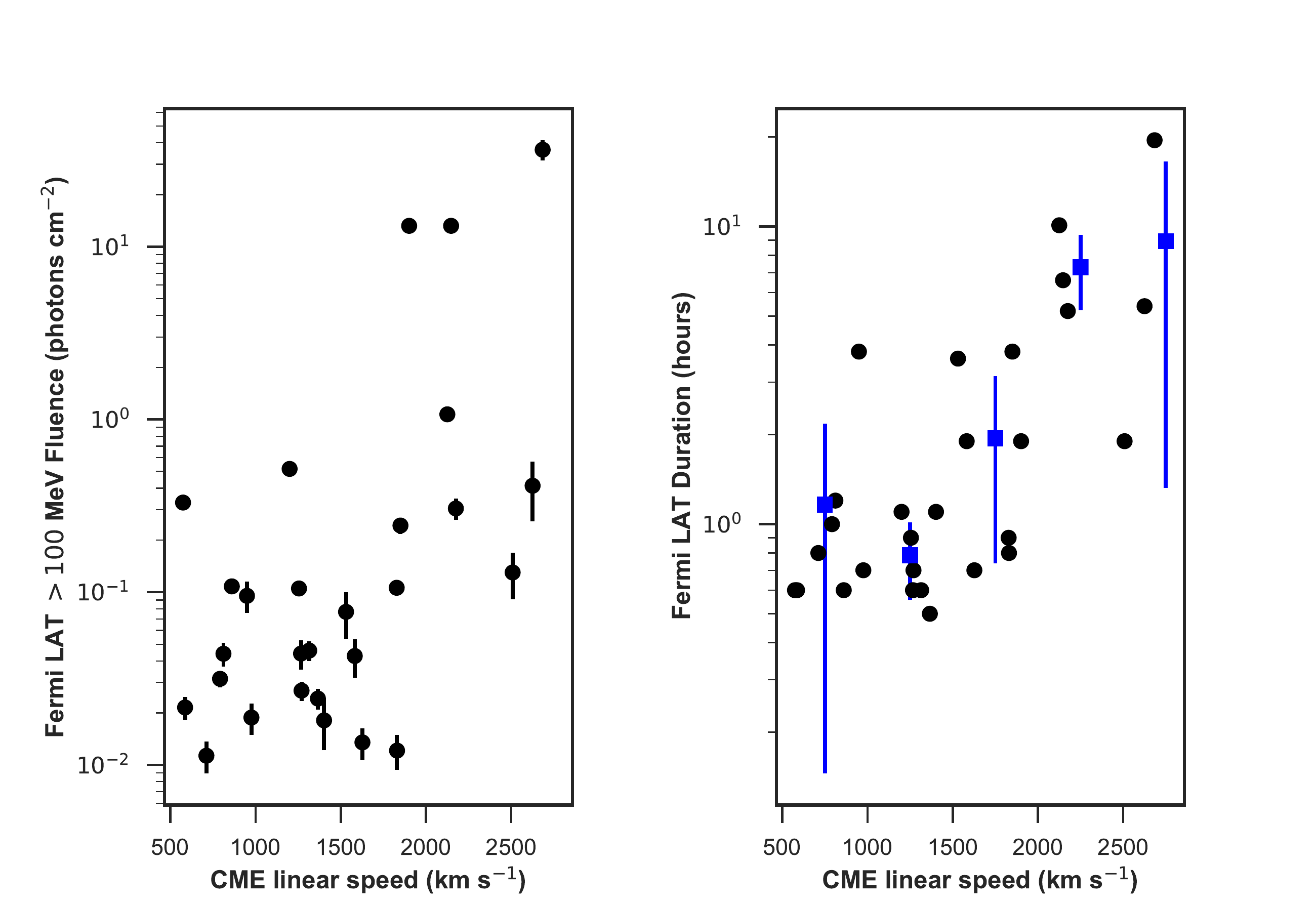}
\caption{\Fermi LAT fluence (from Table 1, left) and duration (right) compared to the associated CME linear speed. There is a trend of higher fluence and longer duration gamma-ray flares being associated with fast CMEs with speeds $> 1500$\,km\,s$^{-1}$. The second panel shows this clearly by including the mean and standard deviation of the flare duration (in blue), in 500\,km\,s$^{-1}$ bins.  
}\label{fig-cmes}
\end{figure}

Figure~\ref{fig-cmes} shows the relationship between the gamma-ray flare fluence and duration as a function of the CME linear speed. In both cases, we find a trend of { higher gamma-ray fluence (Kendall's $\tau$ of 0.46 with p-value of $0.03$) and duration (Kendall's $\tau$ of 0.50 with p-value of $0.02$)} with high CME speeds. This is particularly true for the relationship with gamma-ray flare duration. To illustrate this, the median and standard deviation of duration is shown for bins of 500\,km\,s$^{-1}$. Below a CME speed of 1500\,km\,s$^{-1}$, the median gamma-ray flare duration is 40\,min. The standard deviation is 40 min, but due to one outlying event from 2012-03-09, which occurred close in time to a period of heightened solar activity including a very fast CME two days earlier that is associated with the longest duration gamma-ray flare. Excluding the 2012-03-09 flare, the median and standard deviation is 42 $\pm$ 13 minutes for gamma-ray flares with slow CME speeds. For gamma-ray flares associated with faster CMEs, the average duration increases to 2.2 hours for flares with associated CMEs from 1500-2000\,km\,s$^{-1}$ and $\sim 7$ hours for flares associated with CMEs faster than 2000\,km\,s$^{-1}$. The standard deviations are from 1-5.5 hours, respectively. We note that the \Fermi LAT flare fluence is the integrated flux from time periods where \Fermi was pointed at the Sun and is therefore a lower limit, particularly for the longest duration flares. Similarly, the durations are calculated from the first to last LAT detection, while Fermi was pointed at the Sun. If \Fermi was always pointing towards the Sun, the gamma-ray flare duration and intensity would be more accurate since the start and stop time would be precise. We would then expect less scatter (i.e., lower standard deviations) in these statistics and tighter constraints on CME speed-gamma-ray flare relationships. 

\subsection{Properties of CMEs Near in Time with pre-\Fermi LAT Flares}
To test if past long duration gamma-ray flares from earlier studies were also associated with fast CMEs, we searched for the associated CMEs of the 13 LDGRFs observed between 1982 and 1991, prior to the launch of \Fermi LAT. Unfortunately, no CME catalogs exist which correspond to the 1990 and 1991 gamma-ray events, thus the times of only 5 out of the 13 pre-\Fermi LAT flares could be cross-examined with recorded CME properties. We found CME association with 4 out of 5 of the pre-\Fermi LAT flares from 1982 - 1989. Durations for these 5 gamma-ray events are from 10 to 25 minutes. The associated CME speeds range from 1100 to 1828\,km\,s$^{-1}$. Since the instruments for detection of the pre-LASCO CMEs are different, it is difficult to make a deeper comparison. However, there does appear to be a fast CME close in time with the gamma-ray flares.

\begin{figure}
\centering
\includegraphics[width=0.95\linewidth]{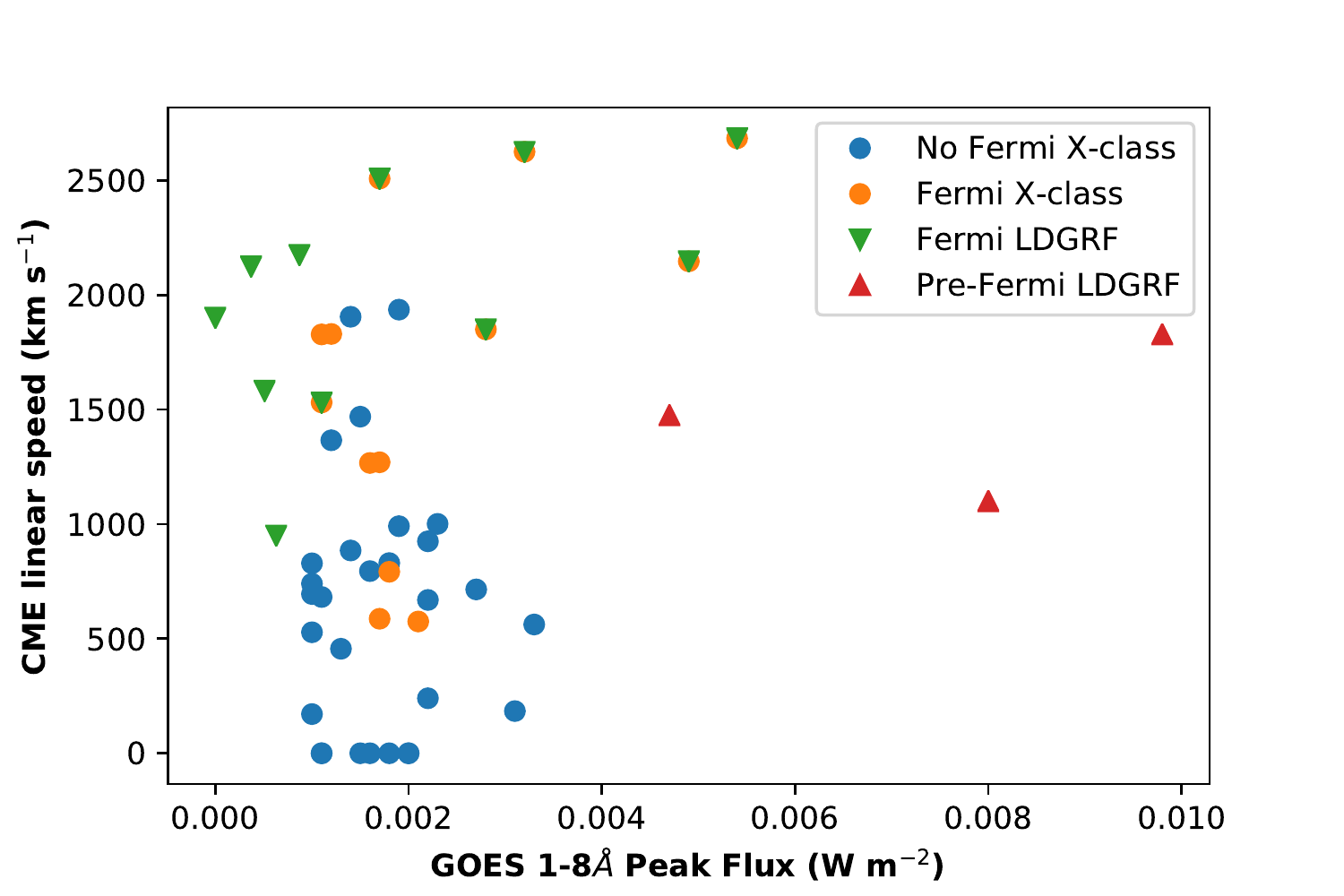}
\caption{Comparison of CME linear speed with X-ray flux for LDGRFs and all GOES X-class flares. LDGRFs are most likely to occur close in time to a fast, wide CME. In the time period where \Fermi was operational, the majority of LDGRFs (green) occur with a median CME speed of 2125\,km\,s$^{-1}$. No LDGRFs are associated with slow CMEs.}\label{fig-cmeXray}
\end{figure}

\subsection{Fast, Wide Angle CMEs and X-class flares with No Fermi LAT Detection}

In our reverse analysis, we examined the CME properties of large solar events where associated gamma-ray emission was not detected in the \texttt{SunMonitor}. First, we searched the SOHO CME list for fast, halo CMEs with velocities above 1500\,km\,s$^{-1}$ to determine whether they were associated with gamma-ray flares below the TS threshold we used to classify gamma-ray flares. There were 4 CMEs, with speeds ranging from $\sim 1600 - 1900$ km\,s$^{-1}$, that we did not already associate with a high signal-to-noise detection. \Fermi was observing the Sun for 2 of these cases. For the 2011-09-22 event, with a CME arrival time observed at 10:48, the Sun was in the field of view of \Fermi from approximately 10:48 to 11:24. For the 2011-09-24 event, with a CME arrival time observed at 12:48, the Sun was in the field of view of \Fermi from approximately 12:15 to 12:55. From a search of the Fermi LAT catalog (the first LAT solar flare catalog is currently in preparation), we found slight increases in the LAT spectra in every case, though they were not significant detections. The details of these events are included in Table~\ref{tbl-fermi}. In both cases, the flare site was located in the eastern hemisphere (N09E89 and N10E56, respectively). It is possible, therefore, that in these cases the gamma-ray emission was primarily located outside our range of view such that a significant detection was not possible. We point out that while the majority of LDGRFs are in the north-eastern hemisphere, they are not as close to the limb as for these two events.  

Second, we compared the associated CME properties for X-class flares with (Table~\ref{tbl-fermi}) and without (Table~\ref{tbl-nofermi}) gamma-ray emission. Figure~\ref{fig-cmeXray} summarizes the result. For low CME speed, there are no LDGRFs detected. Further, gamma-ray emission is not detected in the majority of flares associated with a slow CME or no CME. The median CME linear speed for X-class flares without gamma-ray emission is 768\,km\,s$^{-1}$. For X-class flares with a \Fermi detection, the median speed is 1828\,km\,s$^{-1}$. The CME speed is highest for those with a LDGRF, with a median speed of 2125\,km\,s$^{-1}$. 

Therefore, we find that LDGRFs are associated with fast, CMEs. X-class flares not associated with a fast, CME are unlikely to be associated with gamma-ray emission. Further, no X-class flares without a fast CME are associated with a LDGRF. In the few cases where a fast CME occurred with no \Fermi detection, the locations were towards a solar limb. 

\begin{figure}
\centering
\includegraphics[width=0.5\linewidth,angle=270]{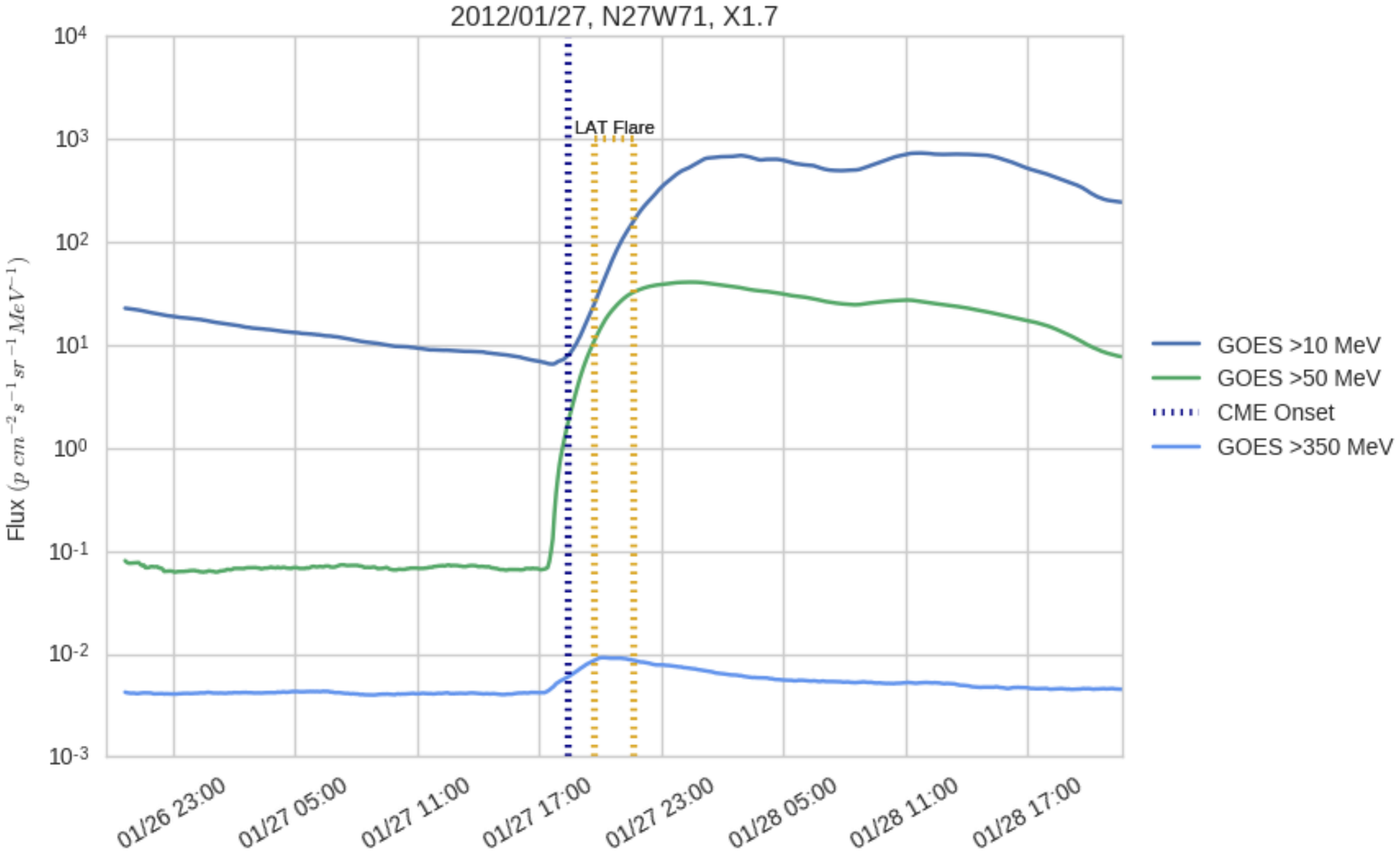}
\includegraphics[width=0.5\linewidth,angle=270]{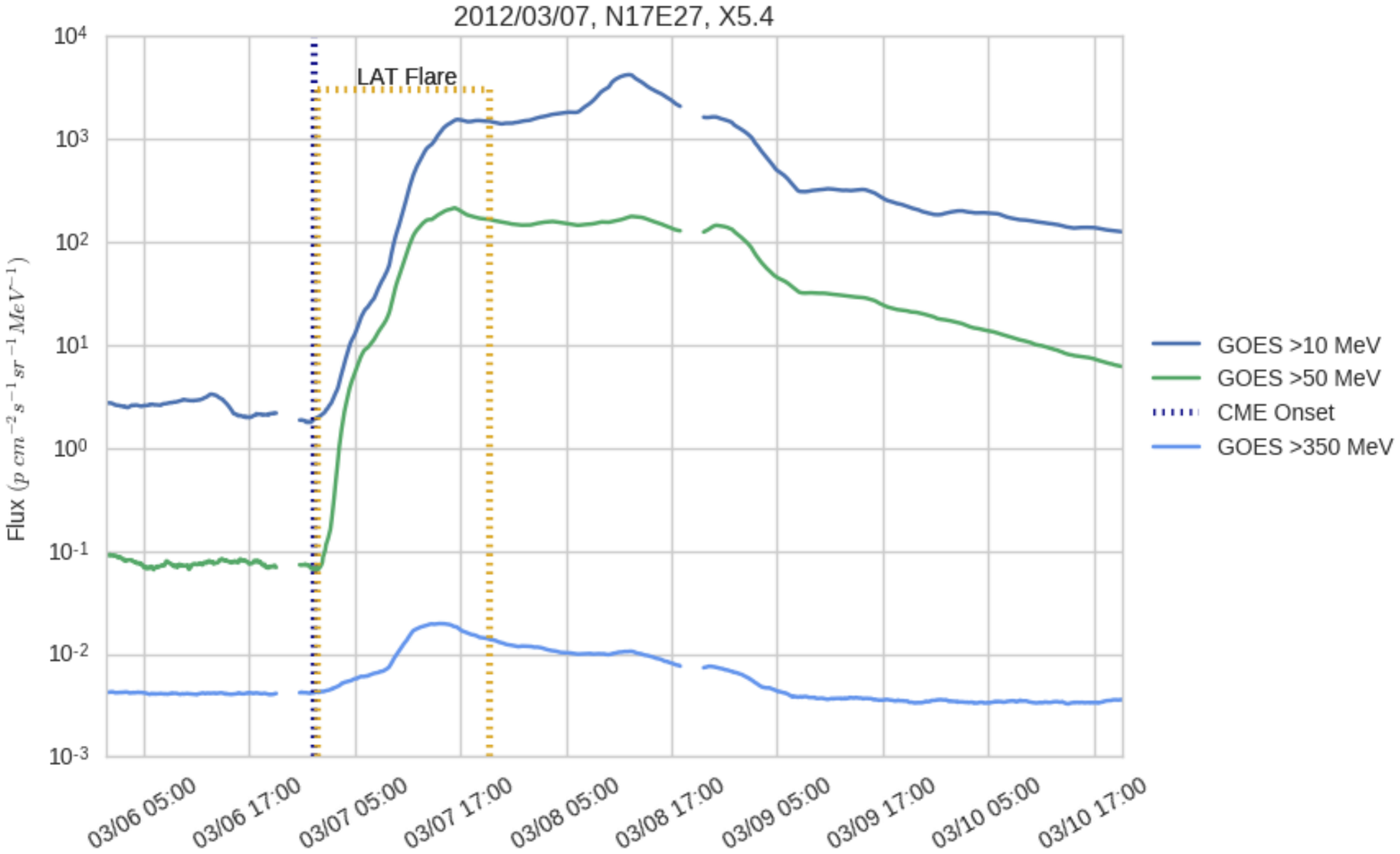}
\includegraphics[width=0.5\linewidth,angle=270]{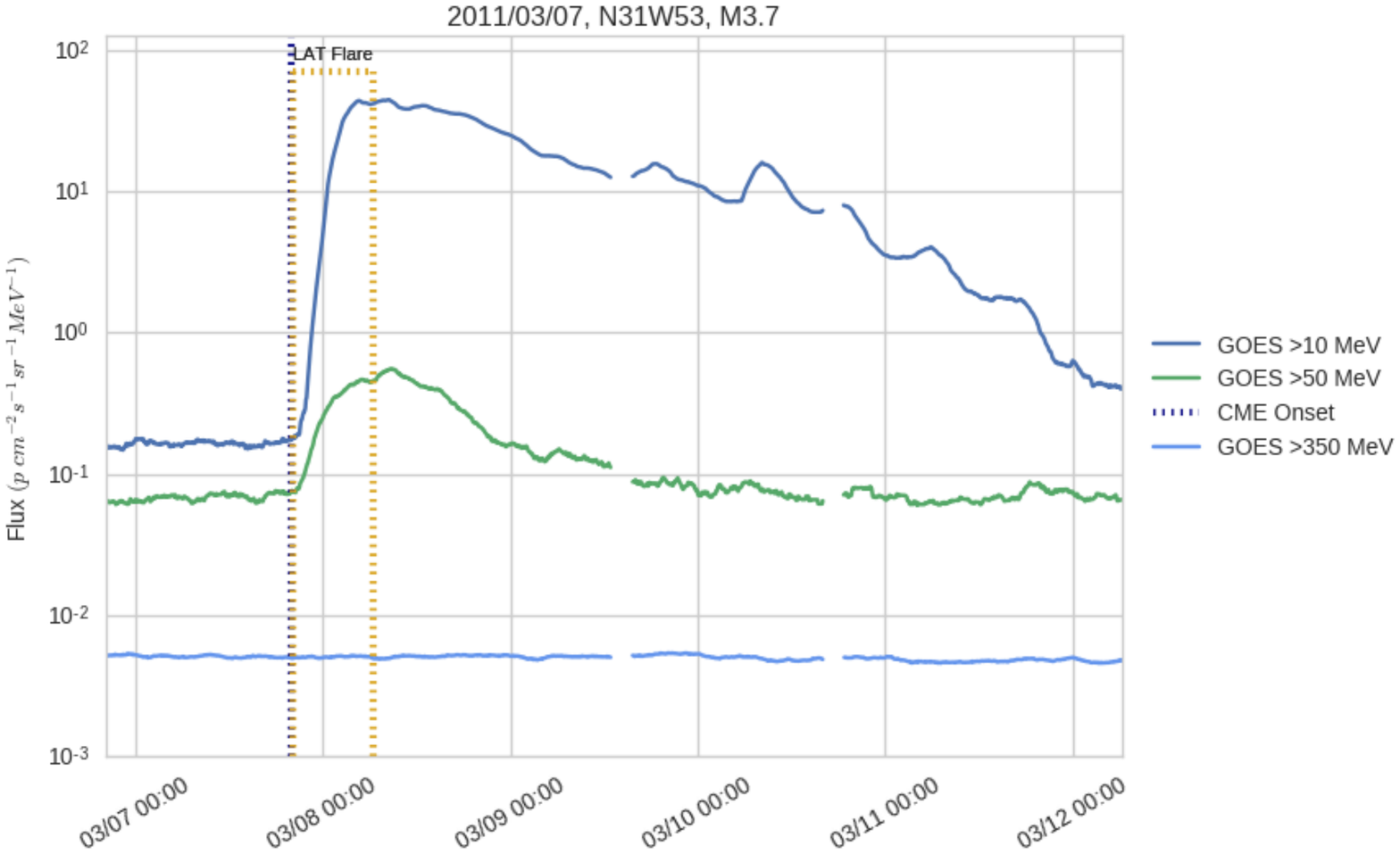}
\caption{
Example GOES proton measurements corresponding with long duration solar gamma-ray flares. The time of the CME onset and \Fermi LAT detection are shown with vertical lines. The HEPAD observations, at energies $> 350$\,MeV, are expected to be detected when the \Fermi flare location is well-connected to the Earth.  In the top two examples, SEP events are detected at all proton energies shown. In the last example (2011/03/07), although the flare is well-located in the western hemisphere, no high energy protons are detected by GOES. We discuss possible explanations in the text.
}\label{fig-seps}
\end{figure}

\section{Solar Energetic Particle Events}
{ SEP events can be categorized as impulsive or gradual (e.g., \citealt{2013SSRv..175...53R}). Impulsive SEP events are considered to accelerate at flare sites at solar longitudes that are magnetically well connected to the observer. Gradual SEP events, which are more intense than impulsive events and last for time periods of hours to weeks, are accelerated in CMEs. SEP transport along a magnetic field can be divided into two modes. The SEP experiences gyration, or spiral motion perpendicular to the magnetic field direction, as well as translation parallel to the magnetic field direction. As a particle'€™s pitch angle (angle between the particle's velocity vector and the magnetic field vector) decreases, the particle experiences faster effective motion along the magnetic field line. Magnetic field lines nearly follow the Sun's radial direction close to the Sun and become more curved with increasing distance from the Sun as the angle between the magnetic field vector and the Sun's radial vector widens. The occurrence and strength of the SEP depends on the magnetic connection between the flare site and the observer's location. In this section, we discuss the properties of SEPs associated with gamma-ray solar flares.
}

\subsection{Properties of SEPs Near in Time with \Fermi LAT Flares}
Gradual solar energetic particle events (SEPs), which are more intense than impulsive events and last for time periods of hours to weeks, are accelerated in CMEs  (e.g., \citealt{2013SSRv..175...53R}). The occurrence and strength of the SEP depends on the magnetic connection between the flare site and the observer's location. If the long duration gamma-ray events are the result of 300 MeV protons accelerated in the shock, we expect that when the acceleration site is well-connected to Earth the GOES satellites will also detect an increase in high energy protons. We investigated this by measuring the occurrence of high energy protons in the P8 channel, through the peak flux when a high energy SEP occurs, along with the time of peak and ratio of the peak flux to the background flux an hour before the flare begins, all shown in Table~\ref{tbl-fermi}. The background flux in the P8 channel was calculated by taking the average of the P8 flux in the time interval from 2 hours before the onset of the SEP event to 1 hour before the SEP onset. We also include the peak $>10$\,MeV proton flux for events associated with a lower energy SEP event. Based on Table~\ref{tbl-fermi}, we find that the peak intensity of the $>10$\,MeV protons is not related to the peak intensity of the higher energy protons. For instance, the highest intensity $>10$\,MeV SEPs on 2012-01-23
and 2012-03-07 are associated with very different P8 peak fluxes of 3.62 and 15.4\,pfu, respectively. This could be due to differences in magnetic connectivity and energy spectra for individual events. We find that the average high energy SEP flux is higher for LDGRFs than { shorter duration gamma-ray flares (i.e., non-LDGRFs)} by 1 pfu, shown in Table~\ref{tbl-stats}. 

Example SEP time profiles are shown for three long duration Fermi LAT flares in Figure~\ref{fig-seps}.The 2012-03-07 flare is the longest duration (14.4 hrs) and highest intensity flare detected by Fermi LAT and is also associated with the highest peak flux in the GOES P8 channel, by an order of magnitude over all but one other high energy SEP event, which occurred 2012-01-23. The gamma-ray flares associated with 300\,MeV SEPs exhibit delay times consistent with the flare location -- i.e., flares in the East have longer delays before the SEP peak than those in the West. 

Half of the \Fermi flares are associated with $> 10$\,MeV SEPs. This is true for both LDGRFs and non-LDGRFs (shown in Table~\ref{tbl-stats}). For the high energy $300$\,MeV protons, we also find enhancements ($F_{P8}/F_{B} > 1.2$) associated with long duration gamma-ray flares in half the events. Of the 4 LDGRFs without an associated SEP, two of these are associated with flares on the eastern limb, which were presumably not well-connected to the Earth. The 2012-03-09 gamma-ray flare occurred during high solar activity and a SEP event was still in progress, meaning the high energy particle background was already high and this should be treated as an LDGRF associated with a SEP. The only flare that is not consistent with the association of gamma-ray flares and 300\,MeV SEPs is the 2011-03-07 event, located at N31W53. For this event, we checked STEREO A and B proton measurements, which showed only weak increases in lower energy particles at each satellite location. This case is a notable counter-example to the origin of LDGRFs from shock acceleration in a CME and therefore warrants further investigation, since the location suggests that the flare site was potentially well-connected and the 10 hour gamma-ray emission was the second longest of any flare. However, this is beyond the scope of the current statistical study.

\subsection{Properties of SEPs Near in Time with pre-\Fermi LAT Flares}
We found similar associations between LDGRFs and SEP properties for the pre-\Fermi LAT solar flare events. At the lower energies, NOAA SWPC records a $> 10$\,MeV proton event near in time to all of the LDGRFs. In cases such as the series of June 1991 flares, the SEP likely results from the combined solar activity occurring during these very active times. 

Based on proton energies $> 10$\,MeV, all of the pre-\Fermi LDGRFs are associated with SEP events (see Table~\ref{tbl-stats}). At the highest energies, we detected increases in the 110-500\,MeV protons (GOES P7 channel) for the majority of LDGRFs. Proton data were publicly available for the events after 1989. In several cases (1989-09-29, 1990-05-24, and 1991-06-11), the enhancement in high energy protons was more than 50 times the background level. All of these events were also associated with very intense X-ray flares of class near or above X10 and flare sites in the western hemisphere. The only LDGRFs without significant enhancements in the high energy protons were those from 1990-04-15 and 1991-03-26. The 1990-04-15 flare site was N32E57 and likely not well-connected with the Earth. The 1991-03-26 flare site was at S28W23 and only a small enhancement (1.2 times the background) was seen. However, this event occurred during a highly active solar time, only 2 days following the most energetic SEP event recorded with a peak $>10$\,MeV proton flux of 43000 pfu on 03-24 at 03:50 UT. It is possible that either there were errors in the GOES measurements following this extreme storm or that the magnetic field configuration was unusual.

To summarize, the pre-\Fermi LDGRFs are associated with SEP events. This is consistent with the connection between CME speed and LDGRFs shown in the previous section. While reliable CME speed measurements are not available for the majority of the pre-\Fermi flares, SEPs are known to be accelerated in CMEs. SEPs are also more likely to be associated with flare sites in the western hemisphere, where the flare site's magnetic connection with Earth is expected to be strongest. Even the LDGRFs in the eastern hemisphere are associated with SEPs, indicating CME particle acceleration that could also be responsible for the production of the observed gamma-ray flares.


\subsection{Energetic SEPs with No Fermi LAT Detection}
Finally, we examined the GOES HEPAD data to determine if any 300\,MeV SEP events occurred that were not associated with a long-duration gamma-ray flare. We conducted an automated search of the HEPAD monthly data between December 2010 and June 2016 to find possible events which met the criteria of having a peak P8 proton flux $> 2.67 \times 10^{-3}$\,pfu and a ratio of the peak P8 proton flux to the background P8 proton flux $>1.38$. These threshold values were chosen as the lowest reasonable peak P8 flux and flux ratio values attributed to high energy protons associated with \Fermi LAT flares lasting $> 1.9$\,hrs, specifically the 2012-03-05 and the 2014-09-01 events. We found three events that had not already been found to be associated with \Fermi LAT detections which met these criteria, which are included at the end of Table~\ref{tbl-fermi}. Of these three events, \Fermi was observing the Sun for only one: the 2013-05-22 event. The Sun was in the field of view of \Fermi from approximately 16:10 to 16:40 for this event, and we found the time of the peak proton flux in the P8 channel to be 16:27. These SEPs without \Fermi LAT detections were not among the strongest energetic particle events, with peak flux at $>10$ MeV from 29-1660 pfu, while the highest intensity SEP during this time period was 6530 pfu. These SEP events were close in time to CMEs, though not necessarily a halo CME (as, for example in the 2012-03-13 event). Two of these events are associated with a flare site towards the western hemisphere limb or over the limb and the last is likely connected to a back side event. It is likely that in the case of the 2013-05-22 event, any potential gamma-ray emission was on the backside of the Sun. Table~\ref{tbl-stats} summarizes the results for associated CME and flare properties of these events relative to the \Fermi LAT detections. The CME speed and X-ray flux associated with these events are lower than that of the properties associated with \Fermi flares and the location is more westward (W68), as is typically the case for SEP events. In particular, since the CME speed and location of these SEP events is lower and westward of the majority of \Fermi flares, it is not surprising that these events did not coincide with gamma-ray flare detections.

\section{Summary}
In this paper, we have investigated the relationship between long duration gamma-ray solar flares and X-ray, CME, and SEP properties to determine whether they originate from proton acceleration via magnetic connection in the flaring region or through the CME shock. Our analysis included a reverse study, in which we further searched for evidence of large solar events (i.e., X-class flares, fast halo CMEs, energetic SEP events) without associated gamma-ray emission. Based on this work, we make a number of conclusions about LDGRFs.

From our analysis of the 45 X-class flares occurring from March 2011 through June 2015, we find that LDGRFs are not necessarily associated with the brightest X-ray flares. In fact, the flare temperature and class are lower than the majority of X-class flares.  Further, we find that there are a number of bright X-class flares which are not associated with gamma-ray emission. 

The main differences between X-class flares with and without a \Fermi detection are flare location and association with a fast, wide CME. LDGRFs, including those detected before \Fermi, are preferentially located in the north-eastern hemisphere. They are also all associated with a fast, wide CME. Particularly, we find that all of the $v > 2000$\,km\,s$^{-1}$ CMEs are associated with a LDGRF and that among the X-class flares without a fast CME, no LDGRFs are detected. The majority of the LDGRFs are also coincident with 300\,MeV SEP events, suggesting that the protons accelerated in the SEP event are related to those producing the gamma-ray flares. A notable counter-example is the 2011-03-07 flare, which was one of the longest LDGRFs and was also well-located, but which did not produce an enhancement in high energy SEPs. Our results lead us to favor the proton precipitation scenario, where the LDGRFs originate from shock acceleration in fast, wide CMEs.

\begin{acknowledgments}
Data used in this analysis are publicly accessible from NASA (Fermi, SOHO/LASCO, Wind/WAVES) and NOAA (GOES). NOAA GOES SEM data were downloaded from here: \url{http://www.ngdc.noaa.gov/stp/satellite/goes/dataaccess.html}.  Details on the satellite and instruments are available in the {\it GOES N Series Data Book} available here: \url{http://satdat.ngdc.noaa.gov/sem/goes/goes\_docs/nop/GOES\_N\_Series\_Databook_rev-D.pdf}.

We thank Ed Cliver (NSO) for valuable discussions and assistance with this work. V. Bernstein gratefully acknowledges funding through NSF and the University of Colorado as part of the LASP 2016 Research Experience for Undergraduates Program in Solar and Space Physics. L. Winter gratefully acknowledges Atmospheric and Environmental Research, who supported much of the work presented. \end{acknowledgments}

\bibliography{Bibliography}

\newpage

\begin{table}
\centering
\caption{Properties of Fermi LAT Gamma-Ray Flares}\label{tbl-fermi}
\begin{tabular}{llllllllllllll}
\hline
\hline
$T_{LAT}$ & D$_{\rm LAT}$ & $F_{LAT}$ & err $F_{LAT}$  & $T_{CME}$ & $v_{CME}$ & $R_{\rm Max}$ & $F_{1-8\AA}$ & Location & $T_{X-ray}$  & $F_{P8}$ & $F_{P8}/F_{B}$  & $T_{P8}$& $F_{\rm SEP}$\\
UT & hrs & ph cm$^{-2}$ & ph cm$^{-2}$ 
& UT & km/s &  & W/m$^2$ & H$\alpha$ & UT &  &   &UT & pfu \\

\hline
2011-03-07 20:15 & 10.1 & 1.065 & 0.088 & 20:00 & 2125 & 0.15 & M3.7 & N31W53 & 19:43 & 2.47 & 1.10 & & \\
\hline
2011-06-02 09:43 & 0.7 & 0.019 & 0.004 & 08:12 & 976 & 0.07 & C2.7 & S19E25 & 09:44 & 2.51 & 1.18 & & \\
\hline
2011-06-07 07:34 & 0.9 & 0.105 & 0.008 & 06:49 & 1255 & 0.15 & M2.5 & S21W54 & 06:49 & 3.67 & 1.35 & 06:19 & 72\\
\hline
2011-08-04 04:59 & 0.6 & 0.046 & 0.006 & 04:12 & 1315 & 0.28 & M9.3 & N19W36 & 03:41 & 1.91 & 1.26 & & 96 \\
\hline
2011-09-06 22:13 & 0.6 & 0.330 & 0.010 & 23:05 & 575 & 0.49 & X2.1 & N14W18 & 22:12 & 2.69 & 1.22 & & \\
\hline
2011-09-07 23:36 & 1.0 & 0.031 & 0.003 & 23:06 & 792 & 0.37 & X1.8 & N14W28 & 22:32 & 2.67 &  1.15 & & \\
\hline
2012-01-23 04:07 & 5.2 & 0.305 & 0.044 & 04:00 & 2175 & 0.28 & M8.7 & N28W21 & 03:38 & 3.62 & 2.01 & 18:13 & 6310 \\
\hline
2012-01-27 19:44 & 1.9 & 0.130 & 0.039 & 18:27 & 2508 & 0.27 & X1.7 & N27W71 & 17:37 & 4.84 & 2.52 & 20:28  & 796\\
\hline
2012-03-05 04:12 & 3.6 & 0.077 & 0.023 & 04:00 & 1531 & 0.27 & X1.1 & N17E52 & 03:17 & 2.67 & 1.13 &  & 35\\
\hline
2012-03-07 00:45 & 19.5 & 36.39 & 4.909 & 00:24 & 2684 & 0.22 & X5.4\tablenotemark{*} & N17E27 & 00:02 & 15.4 & 6.46 & 15:35 & 6530\\
\hline
2012-03-09 05:16 & 3.8 & 0.095 & 0.020 & 04:26 & 950 & 0.28 & M6.3 & N15W03 & 03:22 & 3.12 & 1.01 & & \\
\hline
2012-05-17 02:17 & 1.9 & 0.043 & 0.011 & 01:48 & 1582 & 0.22 & M5.1 & N11W76 & 01:25 & 12.7 & 5.08 &  02:42 & 255 \\
\hline
2012-06-03 17:39 & 0.4 & 0.049 & 0.004 & \nodata & \nodata & 0.19 & M3.3 & N16E33 & 17:48 & 2.32 & 1.19 & & \\
\hline
2012-07-06 23:18 & 0.9 & 0.106 & 0.006 & 23:24 & 1828 & 0.43 & X1.1 & S13W59 & 23:01 & 2.48 & 1.16 & & 25 \\
\hline

2011-08-09\tablenotemark{$\dagger$} & \nodata & \nodata & \nodata &  08:12  & 1610 & 0.56 & X6.9 & N17W69 & 08:05 & 2.67 & 1.41 &  & 26  \\
\hline
2011-09-24\tablenotemark{$\dagger$} & \nodata & \nodata & \nodata & 12:48 & 1915 & 0.30 & M7.1 & N10E56 & 12:33 & 2.87 & 1.22 &  & 35\\
\hline
2012-10-23\tablenotemark{$\dagger$} & \nodata & \nodata & \nodata & \nodata & \nodata &  0.41 & X1.8 & S12E44 & 03:17 & 2.05 & 1.20 \\
\hline
2012-11-27 15:49 & 0.8 & 0.013 & 0.002 & 15:36 & 712 & 0.25 & M1.6 & N05W73& 15:52 & 2.08 & 1.20 & & \\
\hline
2013-04-11 07:01 & 0.6 & 0.108 & 0.007 & 07:24 & 861 & 0.24 & M6.5 & N09E12 & 06:55 & 1.55 & 1.30 & & 114 \\
\hline
2013-05-13 04:32 & 0.7 & 0.027 & 0.003 & 02:00 & 1270 & 0.40 & X1.7 & N11E90 & 01:53 & 2.32 & 1.06 & & \\
\hline
2013-05-13 17:17 & 3.8 & 0.243 & 0.025 & 16:00 & 1850 & 0.41 & X2.8 & N11E85 & 15:48 & 2.21 & 1.24 & & \\
\hline
2013-05-14 01:12 & 5.4 & 0.413 & 0.156 & 01:25 & 2625 & 0.42 & X3.2 & N08E77 & 00:00 & 2.26 & 1.12 & & \\
\hline
2013-10-11 06:32 & 1.1 & 0.517 & 0.017 & 07:24 & 1200 & \nodata & BL & \nodata & \nodata & 1.36 & 1.16 & & \\
\hline
2013-10-25 08:20 & 0.6 & 0.022 & 0.003 & 08:12 & 587 & 0.49 & X1.7 & S08E73 & 07:53 & 1.49 & 1.19 & & \\
\hline
2013-10-28 15:34 & 1.2 & 0.044 & 0.007 & 15:36 & 812 & 0.24 & M4.4 & S06E58 & 15:07 & 1.56 & 1.48 & & \\
\hline
2014-01-06 07:25 & 1.1 & 0.018 & 0.006 & 08:00 & 1402 & \nodata & BL & \nodata & \nodata & 3.91 & 1.84 & 09:51& 42\\
\hline
2014-01-07 18:42 & 0.8 & 0.012 & 0.003 & 18:24 & 1830 & 0.29 & X1.2 & S15W11 & 18:04 &  3.21 & 1.77   & 22:30 & 1033\\
\hline
2014-02-25 01:11 & 6.6 & { 13.15} & 0.409 & 01:25 & 2147 & 0.52 & X4.9 & S12E82 & 00:39 & 2.13 & 1.61 &  20:50 & 103\\
\hline
2014-06-11\tablenotemark{$\dagger$} & \nodata & \nodata & \nodata & 09:24 &  829 & 0.30 & X1.0 & S18E65 & 09:06 & 2.40 & 1.18 \\
\hline
2014-09-01 11:02 & 1.9 & { 13.23} & 0.304 & 11:12 & 1901 & \nodata & BL & \nodata & \nodata & 3.14 & 1.38 &  02/20:17 \\ 
\hline
2014-09-10 17:47 & 0.6 & 0.044 & 0.008 & 18:00 & 1267 & 0.42 & X1.6 & N14E02 & 17:21 & 2.46 & 1.20 & 11/04:48 & 126 \\
\hline
2015-06-21 05:24 & 0.5 & 0.024 & 0.003 & 02:36 & 1366 & 0.26 & M2.0 & N12E16 & 01:02 & 1.63 & 1.27 & & 1070 \\
\hline
2015-06-25 09:25 & 0.7 & 0.014 & 0.003 &  08:36 & 1627 & 0.31 & M7.9 & N09W42 & 08:02 & 1.57 & 1.30 &\\
\hline
\hline
\multicolumn{13}{c}{Fast Halo CMEs Without Fermi LAT Detections}\\
\hline
\hline
\hline
2011-09-22 & \nodata & \nodata & \nodata & 10:48 & 1905 & 0.33 & X1.4 & N09E89 & 10:29 & 2.97 & 1.09 &  & \\
\hline
2012-07-19 & \nodata & \nodata & \nodata & 05:24 & 1631 & 0.24 & M7.7 & S13W88 & 04:17 &  2.59 & 1.02  & \\
\hline
\hline
\hline
\multicolumn{13}{c}{HEPAD SEP Events without Fermi LAT Detections}\\
\hline
\hline
2012-03-13 & \nodata & \nodata &  \nodata & 17:36 & 1884 & 0.31 & M7.9 & N17W66 & 17:12 & 2.78 & 1.59 & 18:17 & 469 \\
\hline
2013-05-22 & \nodata & \nodata &  \nodata & 13:25 & 1466 & 0.16 & M5.0 & N15W70 & 13:08 & 3.34 & 1.74 & 16:27 & 1660 \\
\hline
2015-10-29 & \nodata & \nodata &  \nodata & 01:25 & 390 & & \nodata & \nodata & \nodata & 3.36 & 1.50 & 03:59 & 29 \\
\hline

\end{tabular}
\vspace{0.25cm}

The columns of the table indicate gamma-ray, CME, X-ray, and energetic proton properties associated with the gamma-ray solar flares. These include the onset time of the Fermi LAT flare ($T_{LAT}$), duration in hours of the LAT emission ($D_{LAT}$), 
Fermi LAT total measured fluence with units of photons\,cm$^{-2}$ ($F_{LAT}$), 
SOHO LASCO CME onset time ($T_{CME}$), SOHO LASCO CME linear speed in km\,s$^{-1}$ ($v_{CME}$), $W_{CME}$, maximum ratio of the GOES hard to soft X-ray bands during the flare ($R_{\rm Max} = (F_{0.5-4\AA}/F_{1-8\AA})_{\rm max}$), GOES X-ray class ($F_{1-8\AA}$), H$\alpha$ flare location from the SOHO LASCO CME catalog or the NOAA X-ray flare list, start time of the X-ray flare ($T_{X-ray}$), peak flux in the GOES HEPAD P8 channel with units of $10^{-3}$ pfu ($F_{P8}$), ratio of the peak P8 proton flux to the background P8 proton flux ($F_{P8}/F_{B}$), time of the peak flux in the P8 channel ($T_{P8}$), and the peak $> 10$\,MeV proton flux for NOAA SEP events (F$_{\rm SEP}$) in units of pfu.

\tablenotemark{$\dagger$}{These LAT detections are impulsive detections with the LLE method, described in \citet{0004-637X-787-1-15}. }

\tablenotemark{*}{There is also an X1.3 flare from 01:05-01:23 at N22E12.}

\end{table} 

\newpage

\begin{table}
\centering
\caption{Properties of X-class flares without \Fermi LAT Gamma-Ray Flares}\label{tbl-nofermi}
\begin{tabular}{llllllll}
\hline
\hline
$T_{\rm X-ray}$ &  $T_{CME}$ & $w_{CME}$ & $v_{CME}$ & $R_{\rm Max}$ & $F_{1-8\AA}$ & Location  &  $F_{\rm SEP}$\\
UT & UT & deg & km/s &  & W/m$^2$ & H$\alpha$ & pfu \\
\hline
\hline
2011-02-15 01:56 & 02:24 & 360 & 669 & 0.44 & X2.2 & S20W12  \\
\hline
2011-03-09 23:23 & \nodata & \nodata & \nodata & 0.41 & X1.5 & N09W12 \\
\hline
2011-09-22 11:01 & 10:48 & 360 & 1905 & 0.33 & X1.4 & N09E89 \\
\hline
2011-11-03 20:27 & 23:30 & 360 & 991 & 0.40 & X1.9 & N18E57 \\ 
\hline
2012-07-12 16:49 & 16:48 & 360 & 885 & 0.38 & X1.4 & S15W01 & 96  \\
\hline
2013-05-15 01:48 & 1:48 & 360 & 1366 & 0.32 & X1.2 & N12E64 & 41  \\
\hline
2013-10-28 02:03 & 02:24 & 360 & 695 & 0.32 & X1.0 & N04W66  \\
\hline
2013-10-29 21:54 & 22:00 & 360 & 1001 & 0.56 & X2.3 & N05W89  \\
\hline
2013-11-05 22:12 & 22:36 & 195 & 562 & 0.50 & X3.3 & S13E44  \\
\hline
2013-11-08 04:26 & \nodata & \nodata & \nodata & 0.42 & X1.1 & S14E15  \\
\hline
2013-11-10 05:14 & 05:36 & 262 & 682 & 0.34 & X1.1 & S14W13  \\
\hline
2013-11-19 10:26 & 10:36 & 360 & 740 & 0.41 & X1.0 & S70W14  \\
\hline
2014-03-29 17:48 & 18:12 & 360 & 528 & 0.35 & X1.0 & N11W32  \\
\hline
2014-04-25 00:27 & 00:48 & 296 & 456 & 0.48 &  X1.3 & S15W90  \\
\hline
2014-06-10 11:42 & 11:48 & 111 & 925 &  0.51 & X2.2 & S15E80  \\
\hline
2014-06-10 12:52 & 13:30 & 360 & 1469 &  0.30 & X1.5 & S17E82  \\
\hline
2014-10-19 05:03 & \nodata & \nodata & \nodata & 0.30 & X1.1 & S10E58  \\
\hline
2014-10-22 14:28 & \nodata & \nodata & \nodata & 0.46 & X1.6 & S14E13  \\
\hline
2014-10-24 21:41 & 21:48 & 35 & 184 & 0.44 & X3.1 & S12W21 \\
\hline
2014-10-25 17:08 & 17:36 & 49 &171 & 0.33 & X1.0 & S03E58  \\
\hline
2014-10-26 10:56 & \nodata & \nodata & \nodata & 0.44 & X2.0 & S18W40  \\
\hline
2014-10-27 14:47 & \nodata & \nodata & \nodata & 0.41 & X2.0 & S17W52  \\
\hline
2014-11-07 17:26 & 18:08 & 293 & 795 & 0.29 & X1.6 & N17E40  \\
\hline
2014-12-20 00:28 & 01:25 & 257 & 830 & 0.35 & X1.8 & S21W24  \\
\hline
2015-03-11 16:22 & 17:00 & 74 & 240 & 0.46 & X2.2 & S17E21  \\
\hline
2015-05-05 22:11 & 22:24 & 360 & 715 & 0.38 & X2.7 & N15E79  \\
\hline
\hline
\end{tabular}
\vspace{0.25cm}

The columns of the table indicate CME, X-ray, and energetic proton properties associated with X-class flares with no detected gamma-ray emission. These include the onset time of the X-ray flare ($T_{X-ray}$), 
SOHO LASCO CME onset time ($T_{CME}$), SOHO LASCO CME linear speed in km\,s$^{-1}$ ($v_{CME}$), $W_{CME}$, maximum ratio of the GOES hard to soft X-ray bands during the flare ($R_{\rm Max} = (F_{0.5-4\AA}/F_{1-8\AA})_{\rm max}$), GOES X-ray class ($F_{1-8\AA}$ $\times 10^{-3}$ W m$^{-2}$), H$\alpha$ flare location from the SOHO LASCO CME catalog or the NOAA X-ray flare list, and the peak $> 10$\,MeV proton flux for NOAA SEP events (F$_{\rm SEP}$) in units of pfu. 
\end{table}

\newpage
\begin{table}
\centering
\caption{Properties of Pre-\Fermi LAT Gamma-Ray Flares}\label{tbl-prefermi}
\begin{tabular}{lllllllllll}
\hline
\hline
$T_{\rm GRS/EGRET}$ & D$_{\rm GRS/EGRET}$ & $T_{CME}$ & $v_{CME}$ &  $F_{1-8\AA}$ & Location & $T_{X-ray}$  & $F_{P7}$ & $F_{P7}/F_{B}$  & $T_{P7}$& $F_{\rm SEP}$\\
UT & hrs & UT & km/s  & W/m$^2$ & H$\alpha$ & UT &   & &UT & pfu \\
\hline
\hline
1982-06-03 11:41 & 0.3 & 12:03 & 1100 &  X8.0 & S09E71 & 11:41 & \nodata & \nodata & \nodata & 10 \\
\hline
1984-04-25 & 0.25 & \nodata & \nodata &  X13.0 & S12E43 & 23:56 &  \nodata & \nodata & \nodata & 2500\\
\hline
1988-12-16 08:30 & 0.2 & 08:50 & 1475 & X4.7 & N26E37 & 08:38 & 1.79 & 2.72 & 22:41 & 29 \\
\hline
1989-03-06 13:55 & 0.4 & 14:15 & \nodata &  X15.0 & N35E69 & 13:54 & 0.811 & 1.60 & 11/13:35 & 3500 \\
\hline
1989-09-29 11:33 & 0.3 & 11:27 & 1828 & X9.8 & S26W98 & 10:47 & 271 & 701 & 16:03 & 4500 \\
\hline
1990-04-15 02:44$^{\dagger}$ & 0.5 & \nodata & \nodata &  X1.4 & N32E57 & 02:55 & \nodata & \nodata & \nodata & 12 \\
\hline
1990-05-24 20:47 & 0.1 & \nodata & \nodata &  X9.3 & N36W76 & 20:46 & 22.0 & 56.6 & 21:48 & 180 \\
\hline
1991-03-26 20:27 & 0.2 & \nodata & \nodata &  X4.7 & S28W23 & 20:26 & 0.446  & 1.16 & 27/00:04 & 43000 \\
\hline
1991-06-04 03:39 & 2.8 & \nodata & \nodata &  X12.0 & N30E70 & 03:37 & 2.36 & 6.27 & 05/01:50 & 3000 \\
\hline
1991-06-06 01:07 & 0.3 & \nodata & \nodata &  X12.0 & N33E44 & 00:54 & 3.70 & 9.85 & 07/12:21 & 3000 \\
\hline
1991-06-09 01:36 & 0.25 & \nodata & \nodata &  X10.0 & N34E04 & 01:37 & 1.15 & 3.06 & 18:11 & 3000 \\
\hline
1991-06-11 01:56 & 8.3 & \nodata & \nodata &  X12.0 & N31W17 & 02:09 & 30.4 & 87.6 & 02:17 & 3000 \\
\hline
1991-06-15 08:37 & 1.4 & \nodata & \nodata &  X12.0 & N33W69 & 06:33 & 78.0 & 293 & 10:01 & 1400 \\

\hline

\hline

\hline

\end{tabular}
\vspace{0.25cm}

Gamma-ray, CME, X-ray, and energetic proton properties associated with the early gamma-ray solar flares are shown in the table. Columns include the onset time of the gamma-ray flare ($T_{GRS/EGRET}$), duration in hours of the gamma-ray emission ($D_{GRS/EGRET}$), SOLWIND/SMM CME onset time ($T_{CME}$), SOLWIND/SMM CME linear speed in km\,s$^{-1}$ ($v_{CME}$), $W_{CME}$, GOES X-ray class ($F_{1-8\AA}$), H$\alpha$ flare location from the NOAA X-ray flare list, start time of the X-ray flare ($T_{X-ray}$), peak flux in the GOES EPS P7 channel with units of $10^{-3}$ pfu ($F_{P7}$), ratio of the peak P7 proton flux to the background P7 proton flux ($F_{P7}/F_{B}$), time of the peak flux in the P7 channel ($T_{P7}$), and the peak $> 10$\,MeV proton flux for NOAA SEP events (F$_{\rm SEP}$) in units of pfu. Note: $^{\dagger}$ \citet{1994AIPC..294....3T} reports the start time from Gamma-ray observations with the Soviet GRANAT satellite.

\end{table} 

\newpage
\begin{table}
\centering
\caption{Statistics for CME, Flare, and SEP Properties}\label{tbl-stats}
\begin{tabular}{lllllll}
\hline
\hline
Flare Type & CME speed &  X-ray flux & Location & \% SEP & $F_{P}$ & N \\
\hline
{\bf Gamma-ray Flares} \\
\hline
long duration Fermi & 2011 $\pm 533$ & X2.4 $\pm 1.8$ & N15E27 & 50 & $4.2 \pm 4.2$ & 8\\
long duration pre-Fermi & 1468 $\pm 297$ & X9.5 $\pm 3.7$ & N31E37 & 100 & ($41.1 \pm 80.0$) & 13\\
all Fermi flares & 1482 $\pm 655$ & X1.7 $\pm 0.6$ & N11E27 & 50 & $3.2 \pm 2.9$ & 33 \\
\hline
{\bf Non-GRFs} \\
\hline
X-class flares  & $791 \pm 417$ & X1.6 $\pm 0.7$ & S14E14 & 7.7 & \nodata & 26 \\
CME $> 1500$km/s & 1768 $\pm$ 137 & X1.1 $\pm 0.3$ & S02E01 & 0 & $2.8 \pm 0.2$ & 2\\
SEPs & 1247 $\pm 629$ & M6.5 $\pm 0.2$ & N16W68 & 100 & $3.2 \pm 0.3$ & 3 \\
\hline

\end{tabular}
\vspace{0.25cm}

Statistics are summarized for the average and standard deviation of CME speed, X-ray flare peak flux, flare location, percent of $>10$ MeV SEPs, 300 MeV proton flux in pfu ($F_{P}$), and total number of events for each category. Note that the CME speed for long duration pre-Fermi events is based on only three measurements.
\end{table}

\end{document}